\documentclass[amsmath,superscriptaddress,amssymb,aps,twocolumn,prl]{revtex4}

\usepackage{graphicx}
\usepackage{soul}
\usepackage[colorlinks=true,citecolor=blue,linkcolor=magenta]{hyperref}
\usepackage[usenames]{color}
\usepackage{setspace}

\newcommand{\ket}[1]{\left\vert #1 \right\rangle}
\newcommand{\bra}[1]{\left\langle #1 \right\vert}
\newcommand{\ketbra}[2]{\ket{ #1}\bra{ #2} }
\newcommand{\bla}[1]{\left( #1 \right)}
\newcommand{\blb}[1]{\left[ #1 \right]}

\def \ket#1{|#1\rangle}
\def \bra#1{\langle#1|}

\def \be{\begin{equation}}
\def \ee{\end{equation}}
\def \ba{\begin{array}}
\def \ea{\end{array}}
\def \bea{\begin{eqnarray}}
\def \eea{\end{eqnarray}}

\renewcommand{\phi}{\varphi}

\usepackage{tabularx,ragged2e,booktabs}
\newcolumntype{C}[1]{>{\Centering}m{#1}}

\begin{document}

\title{Signal transduction and conversion with color centers in diamond and piezo-elements}
\author{Jianming Cai}
\affiliation{Institut f\"{u}r Theoretische Physik, Albert-Einstein Allee 11, Universit\"{a}t Ulm, 89069 Ulm, Germany}
\affiliation{Center for Integrated Quantum Science and Technology, Universit\"{a}t Ulm, 89069 Ulm,
Germany}
\author{Fedor Jelezko}
\affiliation{Center for Integrated Quantum Science and Technology, Universit\"{a}t Ulm, 89069 Ulm,
Germany}
\affiliation{Institut f\"{u}r Quantenoptik, Albert-Einstein Allee 11, Universit\"{a}t Ulm, 89069 Ulm, Germany}
\author{Martin B. Plenio}
\affiliation{Institut f\"{u}r Theoretische Physik, Albert-Einstein Allee 11, Universit\"{a}t Ulm, 89069 Ulm, Germany}
\affiliation{Center for Integrated Quantum Science and Technology, Universit\"{a}t Ulm, 89069 Ulm,
Germany}
\date{\today}

\begin{abstract}
The ability to measure weak signals such as pressure, force, electric field, and temperature with nanoscale devices and high spatial resolution offers a wide range of applications in fundamental and applied sciences. Here we present a proposal for a hybrid device composed of thin film layers of diamond with color centers implanted and piezo-active elements for the transduction and measurement of a wide variety of physical signals. The magnetic response of a piezomagnetic layer to an external stress or a stress induced by the change of electric field and temperature is shown to affect significantly the spin properties of nitrogen-vacancy centers in diamond. Under ambient conditions, realistic environmental noise and material imperfections, our detailed numerical studies show that this hybrid device can achieve significant improvements in sensitivity over the pure diamond based approach in combination with nanometer scale spatial resolution. Beyond its applications in quantum sensing the proposed hybrid architecture offers novel possibilities for engineering strong coherent couplings between nanomechanical oscillator and solid state spin qubits.
\end{abstract}

\maketitle

{\it Introduction.---} The sensitive measurement of signals, such as force, pressure, electric field, temperature,
with high spatial resolution possesses a wide range of applications in physics, engineering and the life sciences.
Conventional methods, e.g. for pressure detection can usually operate at length scales down to the micron scale. Going beyond
this regime whilst retaining highest sensitivity represents a significant challenge.

Recently, the negatively charged nitrogen-vacancy (NV) center in diamond \cite{Doherty13} has been established as a
highly sensitive room-temperature nanoscale probe for magnetic \cite{Maze08,Bala08} and electric fields \cite{Dolde11}
as well as temperature \cite{Kucs13,Toyli13,Neumann13} building on the strength of the excellent coherence properties
of an NV-center electron spin. The operational principle of such kind of diamond quantum sensor is based on the fact
that a magnetic/electric field or temperature change will directly affect the spin properties of the NV center which
can then be detected by optical means. The effect of pressure acting on a diamond lattice for example leads to the
coupling between the orbital and spin dynamics of the NV center. This offers a method to measure very high pressure
with a sensitivity better than existing techniques \cite{Doherty13-b}. At room temperature, however, the
coupling between the strain of diamond lattice and the ground state manifold of the NV-center spin is not sufficient
for measuring a low pressure or force.

In this work, we propose a hybrid device that combines a NV-center quantum sensor in diamond with a surface layer of
piezomagnetic material as a signal transducer. The growth of ultrapure isotopically modified diamond and
subsequent nitrogen implantation make it possible to engineer NV centers close to the diamond surface
\cite{Ohno12,Degen12,Stau12,Mueller13} and therefore in close proximity to the piezomagnetic material layer. This
allows for the detection of changes in the stray magnetic field \cite{Bala08,Rondin12,Malet12,Rondin13,Budker11,Budker13}
generated by the piezomagnetic material due to an external stress. We find that the effect of pressure on the NV-center spin
can be enhanced by three orders of magnitude when compared to the effect of the direct application of stress on the
diamond \cite{Doherty13-b}. However, as the NV center lies close to the magnetic material, the magnetic noise is also
expected to become more significant and deteriorate the measurement sensitivity. We analyse the sensitivity of such
a hybrid diamond-piezomagnetic device for the measurement of stress (force) with detailed numerical studies, and show
that a sensitivity of sub-kPa/$\sqrt{\mbox{Hz}}$ (tens of fN/$\sqrt{\mbox{Hz}}$) can be achieved for the measurement
of pressure (force) under ambient conditions with state-of-the-art experimental capabilities by choosing appropriate
control parameters.

In combination with piezoelectric and thermally sensitive elements, our proposed hybrid piezomagentic-NV approach also
provides a new method for the measurement of electric field and temperature with a sensitivity improved by orders of
magnitude as compared to standard diamond NV sensor. The high spatial resolution is achieved thanks to the atomic size of
the NV center and the ability for positioning it with high accuracy \cite{Ohno12,Degen12,Stau12,Mueller13,Erma13,Kucs13,McGuiness11,Hollenberg13}
in nanoscale diamond. Besides its application in signal transduction and precision measurement, the proposed hybrid
system can be exploited to couple the NV-center spin to mechanical oscillations, and significantly enhance the coherent
coupling between these two degrees of freedom. This would allow fast mechanical control of spin qubits, large phonon
induced spin-spin interaction, and efficient phonon cooling by an NV center \cite{MacQuarri13,Hong12,Bennett13,Albrecht13,Rabl09,Kepe13}.

\begin{figure*}
\begin {minipage} {12cm}
\begin{center}
\hspace{-1cm}
\includegraphics[width=6.65cm]{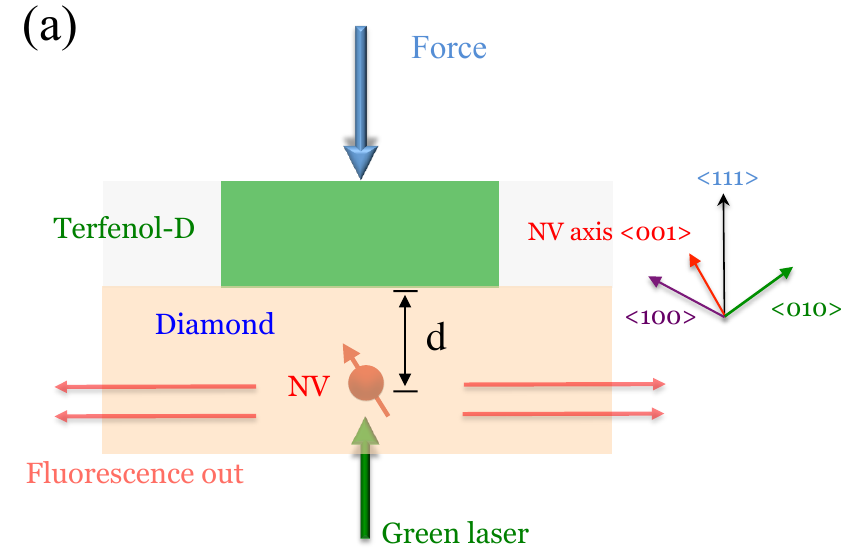}
\hspace{1.5cm}
\includegraphics[width=4.2cm]{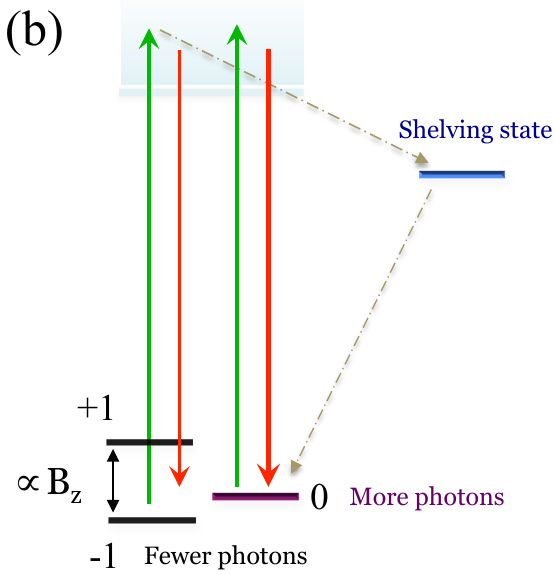}
\end{center}
\end{minipage}
\caption{{\bf A hybrid quantum sensor of a diamond layer (with color centers implanted) and a piezo-active layer.} {\bf (a)} The model hybrid system of diamond (brown) and piezomagnetic film (green). The application of a uniaxial stress induces the change of the magnetostrictive behavior of the piezomagnetic film, which produces a stray magnetic field change measured by the NV center implanted in diamond. {\bf (b)} The energy level structure of a single NV center. The ground spin sublevels ($\pm 1$, $0$) can be initialized and readout via side-collection spin-dependent photoluminescence \cite{Sage12}, and the coherent control of the NV spin is implemented with a microwave field.}\label{fig:setup}
\end{figure*}

{\it The model.---} We consider a system (see Fig.\ref{fig:setup}(a)) made up of a diamond layer grown on the surface of a thin piezomagnetic film such as $\mbox{Tb}_{0.27}\mbox{Dy}_{0.73} \mbox{Fe}_2$ (Terfenol-D) which exhibits a high magnetostrictive effect. The surface area of the piezomagnetic film can be reduced to a designed size by subsequent selective etching. The NV center is implanted in the diamond layer with a depth denoted as $d$ from the interface between the magnetic film and diamond. The ground $^3 A_2$ level of the NV-center spin exhibits a zero field splitting of $D = 2.87 \mbox{GHz}$ between the $m_s=0$ and $m_s=\pm 1$ spin sub-levels. The spin Hamiltonian, including the strain effect of the diamond lattice and the Zeeman interaction with an external magnetic field, is given by
\begin{equation}
    H = D S_z^2 + E(S_x^2-S_y^2) + \gamma \bla{ B_{\parallel}S_z+B_{\perp}S_x} \label{eq:NV_Ham}
\end{equation}
where $\gamma$ is the electron gyromagnetic ratio, $B_{\parallel}$ and $B_{\perp}$ are the magnetic field component in parallel and perpendicular to the NV axis arising from both the applied external magnetic field and the stray magnetic field generated by the piezomagnetic film. The effect of strain on the NV center, as quantified by $E$, induces ground state spin sub-level mixing and is usually much smaller (on the order of $\mbox{MHz}$) than the energy splitting along the NV axis. The zero-field splitting $D$ is slightly affected by the stress due to pressure $\sigma$, with a gradient of about $dD(\sigma)/d\sigma \sim 15 \mbox{kHz} /\mbox{MPa}$ at room temperature \cite{Doherty13-b}. In this work, we will be mostly interested in the case of a weak (sub $\mbox{MPa}$) pressure, and therefore such a direct effect of stress in the diamond lattice on the NV spin is negligible. The coupling strength between an electric field and the NV ground state spin is also negligible in the present model \cite{Oort90}. The spin-dependent  fluorescence of an NV center, see Fig.\ref{fig:setup}(b), provides an efficient mechanism to optically initialize and readout the ground state spin, and to perform optically detected magnetic resonance (ODMR) measurements in the ground state. The resonance frequencies $\omega_{\pm 1}$ in the ODMR spectra correspond to the electronic transitions from the spin sub-level $m_s=0$ to $m_s=\pm 1$ respectively, which depend on the magnetic field acting on the NV center.

The piezomagnetic material consists of magnetic domains, in which the magnetization direction is uniform $\mathbf{M}_{k}=M_S \mathbf{m}_{k}$, where $M_S $ is the saturation magnetization of the material, and $\mathbf{m}_{\alpha}=(\beta_1,\beta_2,\beta_3)$ with $\beta_i (i=1,2,3)$ being the direction cosines with respect to the crystal axes. The response of the piezomagnetic materials that have large magnetostriction, such as Terfenol-D, lead to a change of the magnetization directions of domains, and in turn of the stray magnetic field \cite{Hesjeda05,Norpoth08} that affects the spin properties of NV centers.

\begin{figure*}
\begin{center}
\begin {minipage} {16cm}
\hspace{-1cm}
\includegraphics[width=7.8cm]{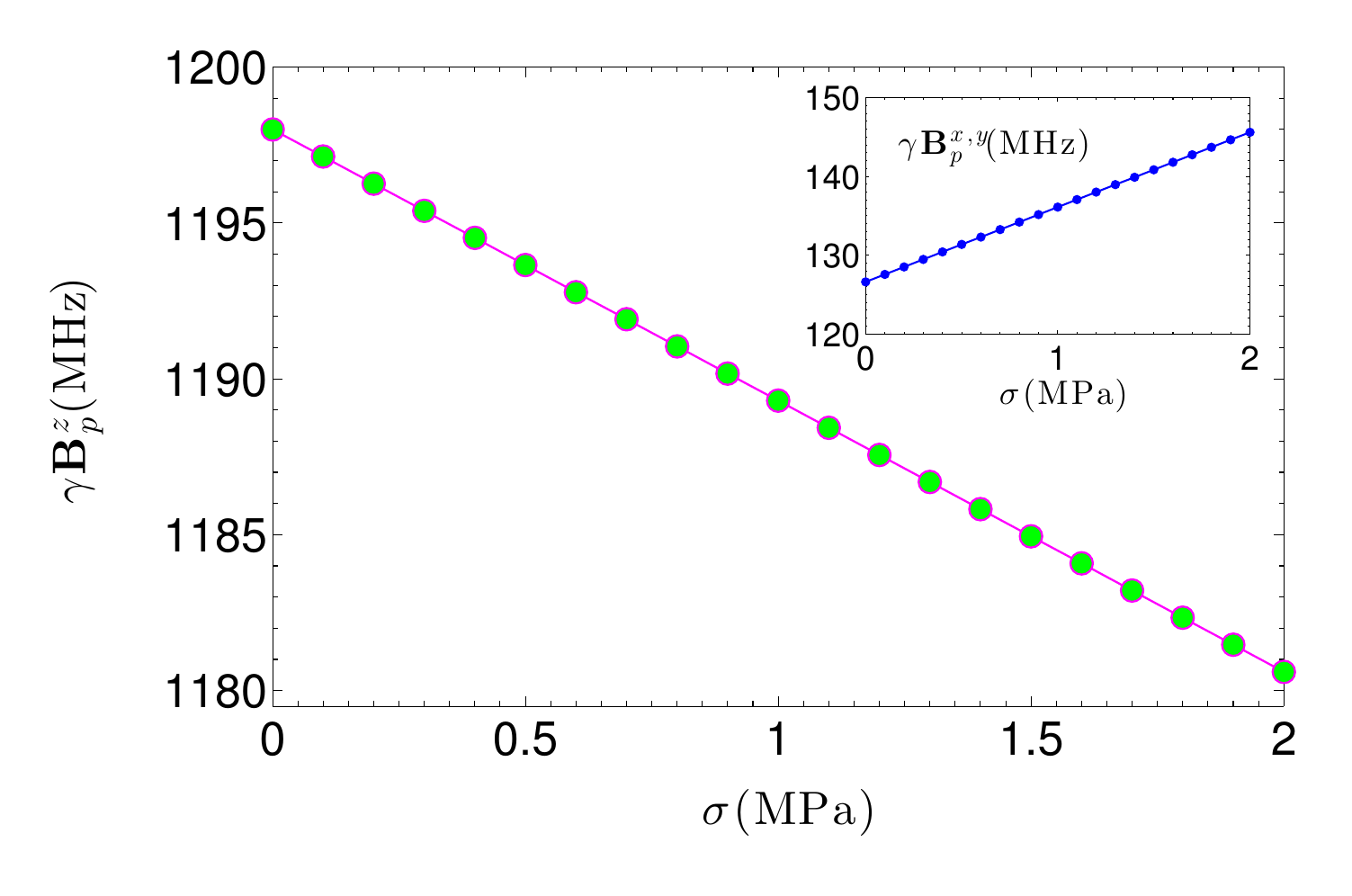}
\hspace{-0.3cm}
\includegraphics[width=7.8cm]{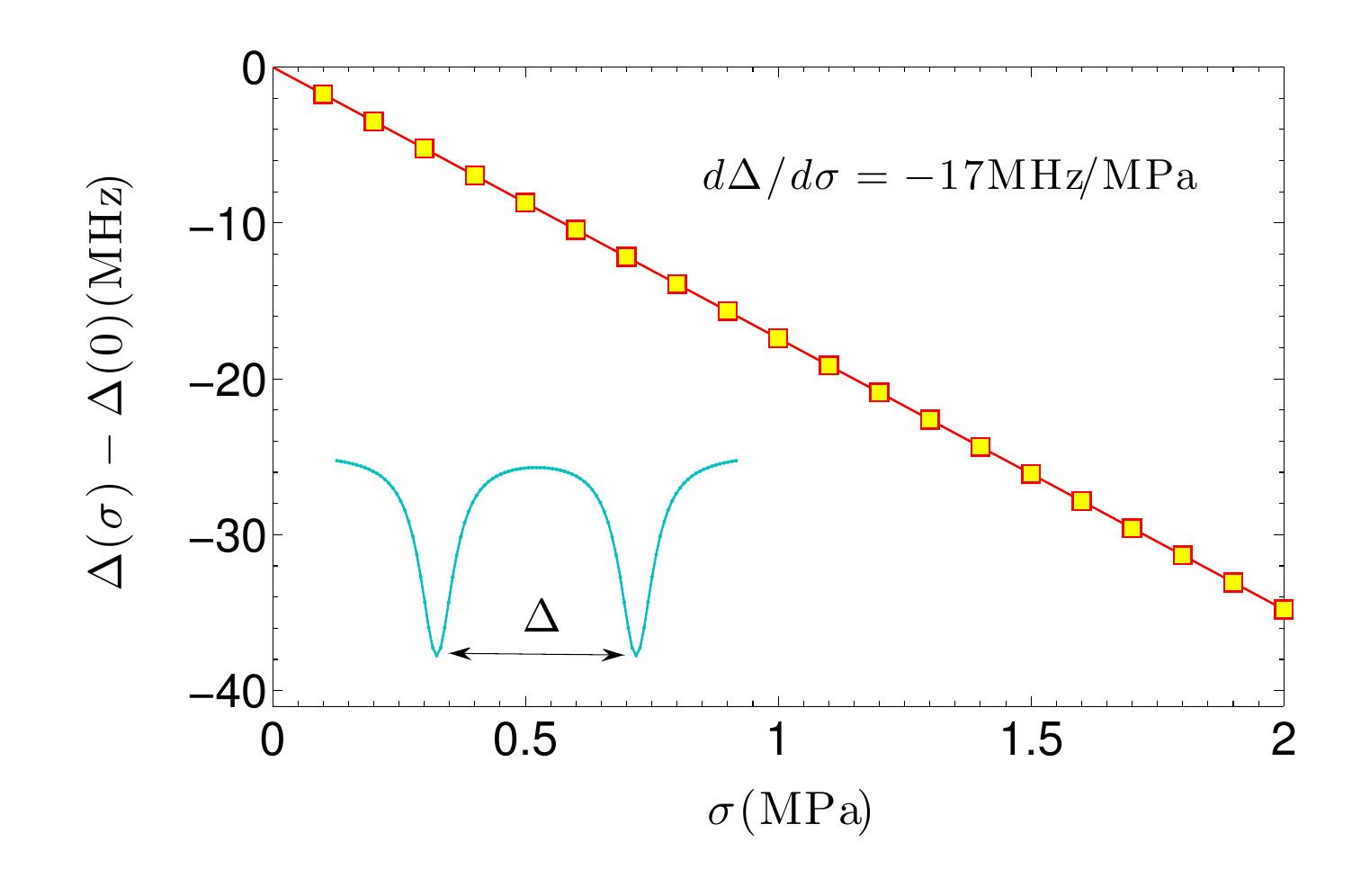}
\end{minipage}
\caption{{\bf The response of the NV center ODMR spectra to the application of a uniaxial stress along the $\langle 111\rangle$ direction at room temperature $\mbox{T=300K}$.} {\bf (a)} The stray field component parallel ($B_{\mbox{z}}$) and perpendicular ($B_{\mbox{x,y}}$, inset) to the NV axis. {\bf (b)} The separation between two resonance frequencies in the ODMR spectra as a function of the pressure $\sigma$. The gradient is estimated to be $d\Delta/d\sigma=-17\mbox{MHz}/\mbox{MPa}$. The dimension of the Terfenol-D film is chosen as $\mbox{(15nm)}^3$, and the distance from the NV to the interface is $d=$15nm. The applied magnetic field is $2350\mbox{G}$ along the $\langle 001\rangle$ direction.}\label{fig:stray_fields}
\end{center}
\end{figure*}

{\it ODMR response to a uniaxial stress.---} To estimate the stress-induced change of the stray magnetic field and the corresponding magnetic noise acting on the NV-center spin, we perform micromagnetic simulation using Landau-Lifshitz-Gilbert (LLG) equation. We first calculate the steady state of Terfenol-D film (characterized by the parameters in Table I of SI) under varying stress by using the LLG micromagnetic dynamic equation (see Eq.1 of SI). To account for the thermal fluctuation at a finite temperature, we add a random thermal fluctuation field $\mathbf{H}_{th}$ to the effective magnetic field $\mathbf{H}_{k}$ in the micromagnetic dynamic equation, and describe the thermal field $\mathbf{H}_{th}$ as a Gaussian random process with the strength determined by the fluctuation-dissipation theorem \cite{Brown63,Brown79,Lazaro98} $\mathbf{H}_{th}^{m,i}(t)=\kappa^{m,i}(t) \bla{\frac{2\alpha k_B T}{\gamma\mu_0 M_S \Delta x^3\Delta t}}^{1/2}$, where $\kappa^{m,i}(t)$ is a normal distributed Gaussian random number that is uncorrelated in time and space while $k_B$ is the Boltzmann constant, $\Delta x$ is the length scale of the computation cell, and $\Delta t$ is the simulation time step, see Supplementary.We apply an external magnetic field along the NV axis, which is assumed to be in the $\langle  001 \rangle$ direction and set it as the $z$ axis, and consider a stress along the $\langle  111 \rangle$ direction with respect to the crystal axes in the piezomagnetic material, along which it has the largest magnetostriction coefficient, see Supplementary Table I.  By solving the Langevin equation using the stochastic Heun scheme \cite{Kampen81}, we calculate the effective stray magnetic field acting on the NV center, and observe a prominent resonance frequency shift in the ODMR spectra, see Fig.\ref{fig:stray_fields}, the shift of $\Delta=\omega_{+1}-\omega_{-1}$ is approximately linear with the stress, and the gradient is about $d\Delta/d\sigma\sim 17 \mbox{MHz}/\mbox{MPa}$ for $d=15$ nm, which is about three orders of magnitude larger than the shift of the zero-field splitting parameter $D$ under a direct pressure \cite{Doherty13-b}. The enhancement would be more significant for shallower NV, e.g. we find  $d\Delta/d\sigma\sim 36 \mbox{MHz}/\mbox{MPa}$ for $d=10$ nm.  We remark that the stress-induced frequency shift of the NV center can be further improved by optimizing the dimension of the piezomagnetic film and the depth of the NV center. The shot-noise-limited sensitivity is characterized by the coherence time $T_2$ that is hindered by the magnetic noise, which we will determine in the following section.

\begin{figure*}
\begin{center}
\begin{minipage}{18cm}
\hspace{-0.45cm}
\includegraphics[width=7.8cm]{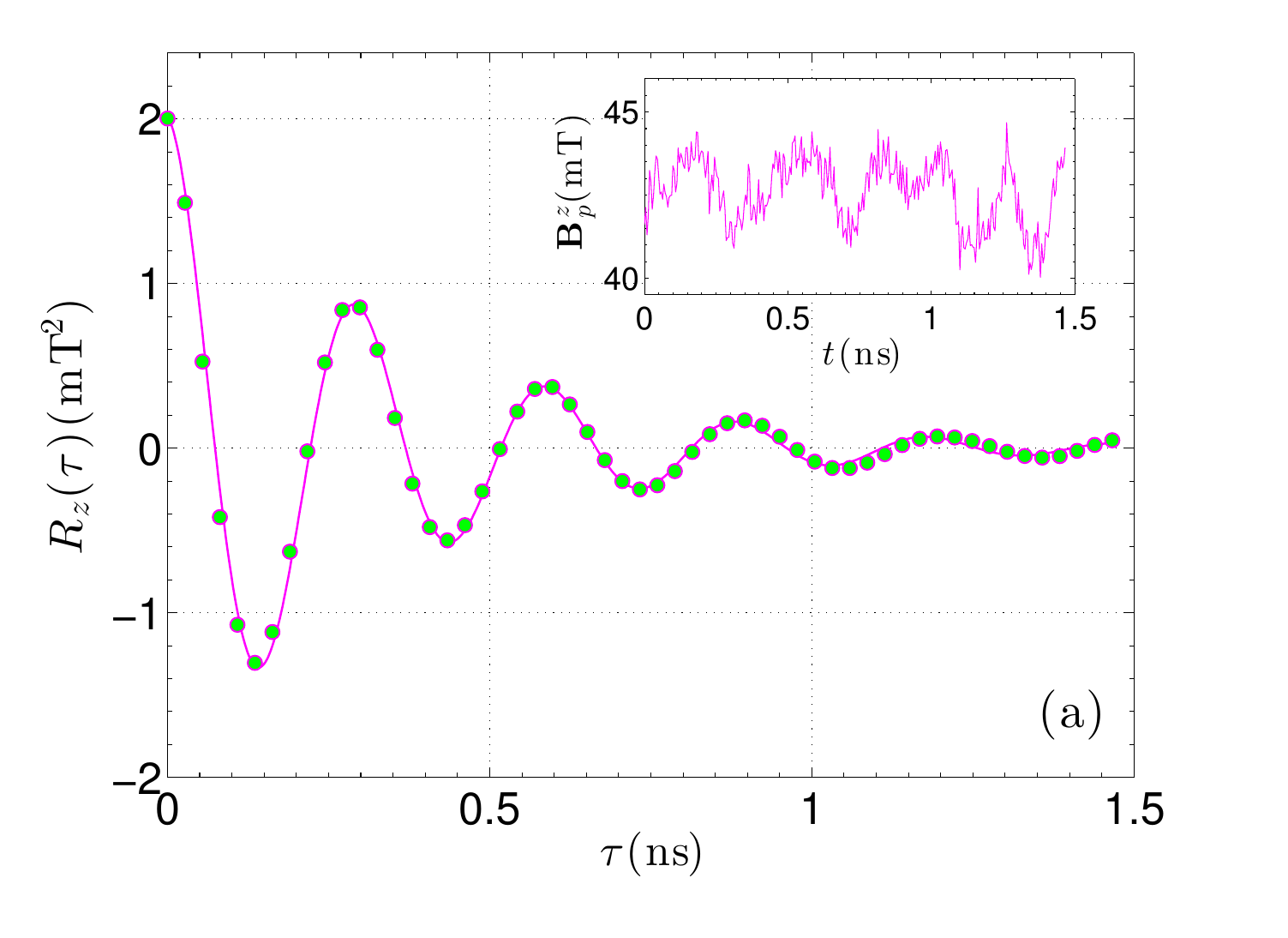}
\hspace{-0.3cm}
\includegraphics[width=7.8cm]{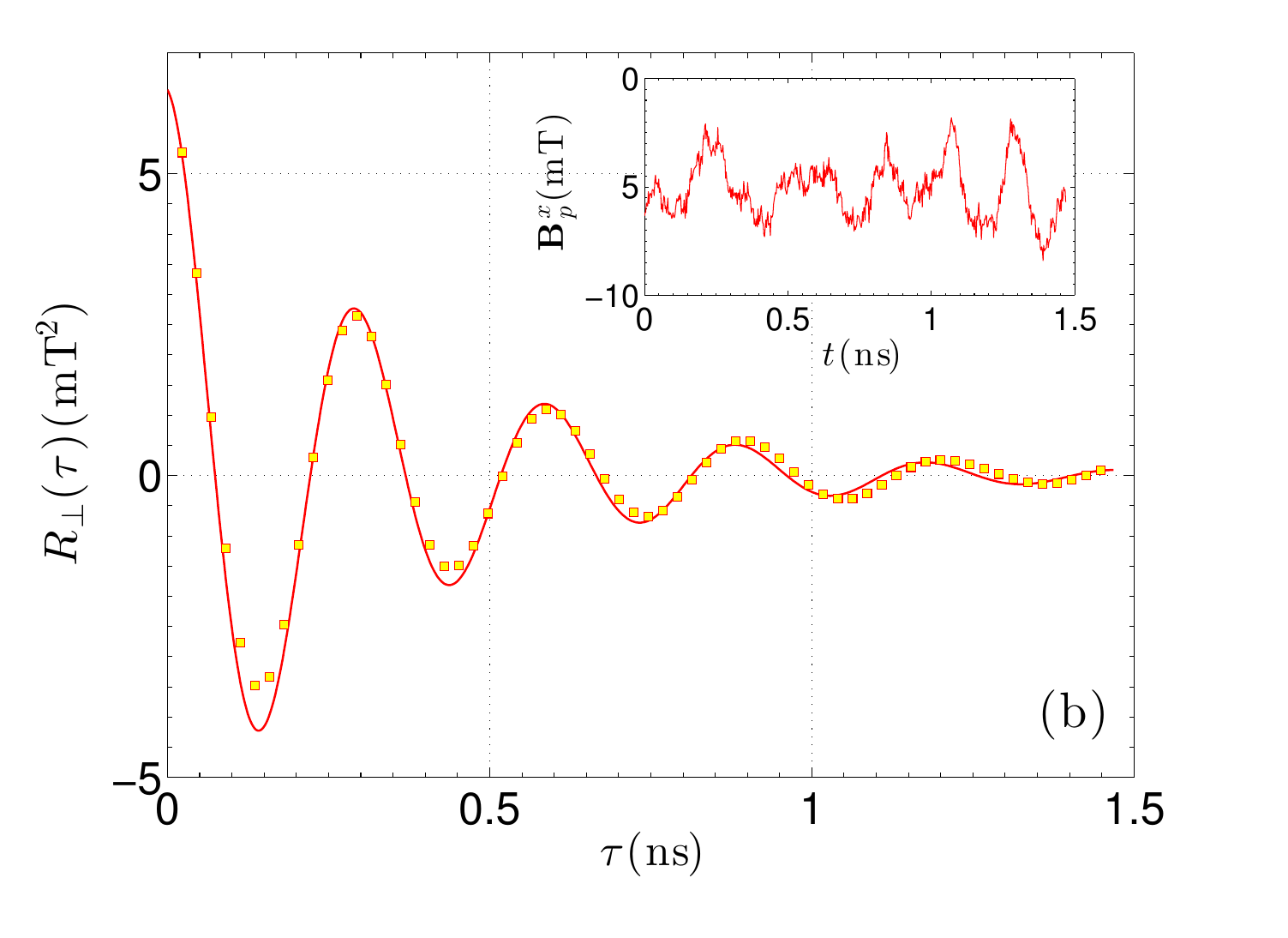}
\end{minipage}
\begin{minipage}{18cm}
\hspace{-0.4cm}
\includegraphics[width=7.8cm]{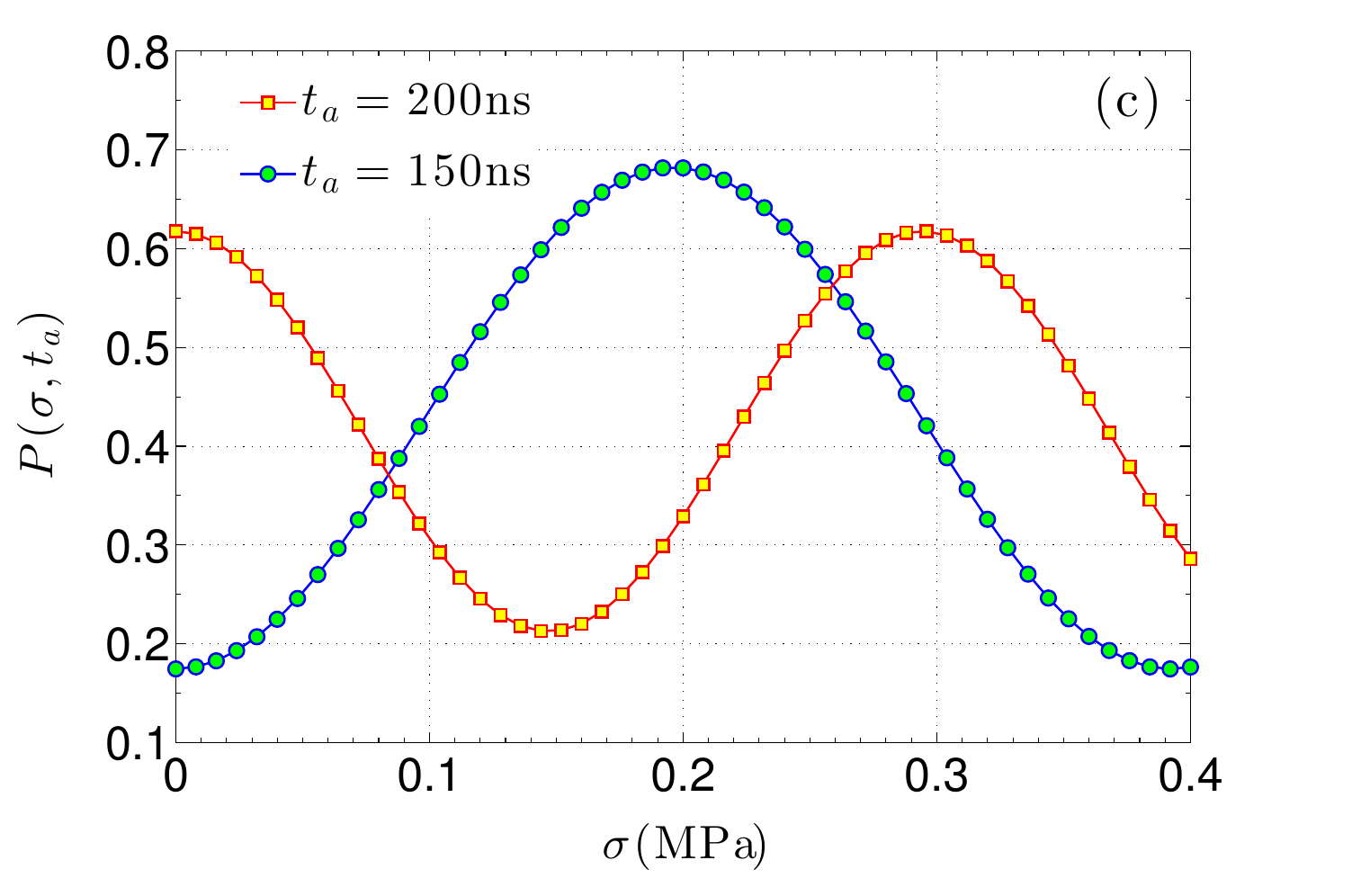}
\hspace{-0.3cm}
\includegraphics[width=7.8cm]{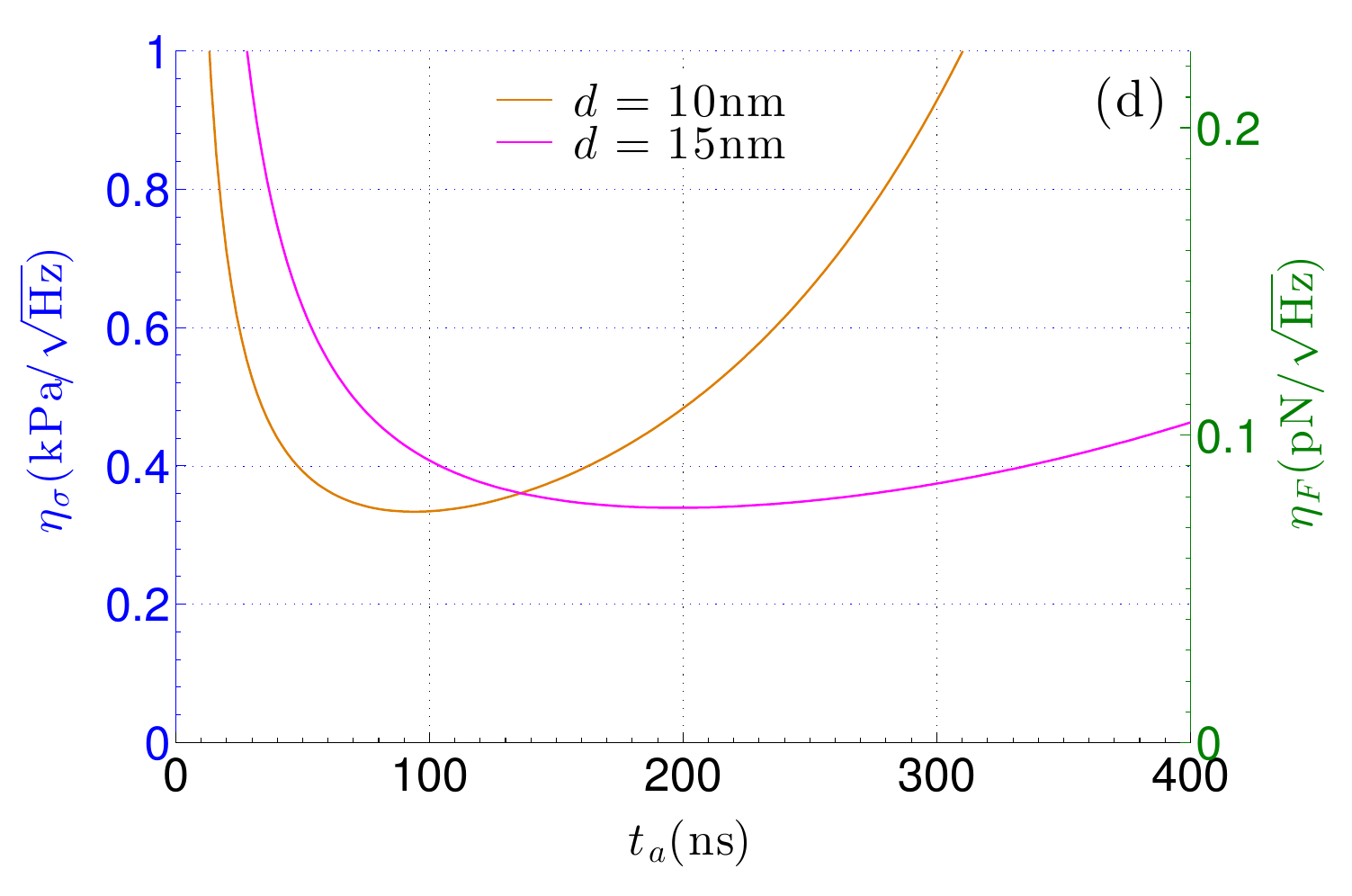}
\end{minipage}
\end{center}
\caption{{\bf Properties of the magnetic noise and the measurement sensitivity of our hybrid system at room temperature $\mbox{T=300K}$}. {\bf (a-b)} The time correlation function of the magnetic noise (parallel (a) and perpendicular (b) to the NV axis) emanating from the magnetic material resulting in a Lorentzian form with the central frequency at $\pm\omega_0=\pm(2\pi)3.38\mbox{GHz}$ and the linewidth $\xi=2.8\mbox{ns}^{-1}$, see Eq.\ref{eq:noise_spec} in the text. The inset shows one random trajectory of the stray magnetic field acting on the NV-center spin produced by the thermal fluctuation in the magnetic material. {\bf (c)} Example of measured signal as a function of the applied uniaxial stress $\sigma$ at the interrogation time $t_a$. The signal contrast decreases for the larger $t_a$ due to the decoherence caused by the magnetic noise. {\bf (d)} The shot-noise-limited sensitivity for the measurement of stress (see Eq.\ref{eq:sens}) and force as a function of interrogation time. The value of $\mathcal{C}$ is $0.3$, the NV spin preparation and readout time is $t_p=600 \mbox{ns}$. The other parameters of the hybrid system are the same as in Fig.\ref{fig:stray_fields}.} \label{fig:noise_spec}
\end{figure*}

{\it Noise power spectrum and measurement sensitivity.---} At a finite temperature, the thermal fluctuation of the piezomagnetic film induces a fluctuating stray magnetic field acting on the NV-center spin, which gives rise to the ODMR linewidth broadening and thereby the decay of spin coherence. Using a pulsed sensing scheme, the NV spin is first prepared into a coherent superposition state $\ket{\phi}=\sqrt{\frac{1}{2}}\bla{\ket{{-1}}+\ket{{+1}}}$ by applying a microwave driving field $H_d=\Omega\blb{ \cos(\omega_{+1}  t)\ketbra{{+1}}{0}+\cos(\omega_{-1} t)\ketbra{{-1}}{0}}+h.c.$ for a pulse duration $t_{\pi/2}=\pi/\Omega $. We remark that by choosing the spin sub-levels $m_s=-1$ and $+1$ to encode quantum coherence, the temperature instability \cite{Kucs13,Toyli13,Neumann13} will shift the energies of these two levels equally, and thus would not induce dephasing. The ideal evolution of the NV spin is $\ket{\phi(t)}=\sqrt{\frac{1}{2}}\bla{\ket{{-1}}+e^{-it\Delta}\ket{{+1}}}$. The coherence time of an NV center in isotopically engineered diamond can exceed 100 $\mu s$ \cite{Bala09,Zhao12}, thus the dominant magnetic noise in the present model arises from the fluctuation in the magnetic material. By solving the stochastic Langevin equation including the thermal random field to the simulation of micromagnetic dynamics, we obtain the time correlation $R_{\zeta}(t)=\langle B_{\zeta}(t) B_{\zeta}(0)\rangle-\langle B_{\zeta} (t) \rangle\langle B_{\zeta}(0)\rangle$ of the magnetic noise from the Terfenol-D film, where $\zeta=x,y,z$ represent the field component along the $x,y,z$ direction. In Fig.\ref{fig:noise_spec}(a), we plot the noise correlation function, which shows a fast decay on a time scale on the order of nanosecond. The correlation time of the magnetic noise, i.e. the decay time scale of the noise correlation function, is dependent on the Gilbert damping constant $\alpha$ (for Terfenol-D, the value of $\alpha\sim 0.1$ \cite{Walo08}). The noise correlation function can be well fitted  by $R_{\zeta}(t)=R_{\zeta}(0)e^{-t\xi}\cos(\omega_0 t )$, and the corresponding power spectra is obtained as follows
\begin{equation}
    \mathcal{S}_{\zeta}(\omega)=\frac{R_{\zeta}(0)}{2}\blb{\frac{\xi}{\xi^2+\bla{
    \omega-\omega_0}^2}+\frac{\xi}{\xi^2+\bla{ \omega+\omega_0}^2}}\label{eq:noise_spec}.
\end{equation}
where $1/\xi$ is the magnetic noise correlation time.

As the correlation time of the magnetic noise (nanosecond) is much shorter than the coherence time of the NV spin, we can describe the NV spin dynamics under the environmental noise by a quantum master equation as derived in supplementary information. The fluorescence measurement of the state $m_s=0$ population after an interrogation time (namely free evolution) $t_a$ and a subsequent $(\pi/2)$ pulse is given by
\begin{equation}
    P(\sigma,t_a)=\frac{1}{8}\blb{3+\chi_{\perp}^2(t_a)}+\frac{1}{2}\cos\blb{2\pi\Delta\bla{\sigma} t_a} \chi_{\perp}(t_a) \chi_{\parallel}(t_a).
\end{equation}
where the dephasing and relaxation coefficients $\chi_{\parallel}(t)=\exp{(-4t \mathcal{S}_{z}(0))}$, $\chi_{\perp}(t)=\exp{(-t \mathcal{S}_{\perp}(\omega_{-1})/2)}$ are determined by the magnetic noise power spectrum, where $ \mathcal{S}_{\perp}=( \mathcal{S}_{x}+ \mathcal{S}_{y})/2$.

To obtain the optimal measurement sensitivity, we choose $t_a$ around the value such that the signal has the maximum absolute slope with respect to the stress, namely $|\sin(t_a\Delta )|=1$. The shot-noise-limit sensitivity for the measurement of stress within a total experiment time $\tau$ is obtained as follows
\begin{equation}
    \eta_{\sigma,t_a,\tau}=\frac{\sqrt{(3+\chi_{\perp}^2(t_a))(5-\chi_{\perp}^2(t_a))}}{8\pi\mathcal{C}
    \chi_{\perp}(t_a) \chi_{\parallel}(t_a) t_a \bla{\frac{d\Delta}{d\sigma}}}\sqrt{\frac{t_a+t_p}{\tau}} \label{eq:sens}
\end{equation}
where $t_p$ is the preparation and readout time in each experiment run, $\mathcal{C}$ accounts for the NV state initialization, readout efficiency, and the signal contrast. The value of $\mathcal{C}$ can exceed $0.3$ by improving the collection efficiency using a plasmonic or optical waveguide \cite{Sage12,Chang06}. The quantity $\eta_{\sigma,t_a,\tau}$ characterizes the smallest change in the stress that can be distinguished within the total measurement time $\tau$. By choosing a larger interrogation time of $t_a$, it is possible to improve the measurement sensitivity, while in the mean time the NV-center spin will suffer more from the magnetic noise. This implies an optimal interrogation time to achieve the best sensitivity. In Fig.\ref{fig:noise_spec} (c-d), we plot the fluorescence signal at two different interrogation times, and the shot-noise-limit sensitivity for the measurement of stress (force), which can achieve $\eta_{\sigma}\sim 0.35 \mbox{kPa}/\sqrt{\mbox{Hz}}$ and $\eta_{F}\sim 75 \mbox{fN}/\sqrt{\mbox{Hz}}$ respectively. We remark that the sensitivity may be further improved by optimizing the dimension of the hybrid system, and using an array of NV centers.

{\it Electric field and temperature measurement.---} Although the NV center has been used to measurement
electric field \cite{Dolde11} and temperature \cite{Kucs13,Toyli13,Neumann13} while the sensitivity is
largely limited by the efficiency of the direct coupling between the NV-center spin and electric field
(temperature). Here, we propose to combine the hybrid diamond and piezomagnetic device with a piezoelectric
element island on a substrate to measures electric field (see supplementary information). An electric
field $\mathrm{E}$ induces a strain $\epsilon=\epsilon_e\cdot \mathrm{E}$ of the piezoelectric element
island, which generates a stress $\sigma=\epsilon\cdot Y$ acting on the attached piezomagnetic layer
where $Y$ denotes the Young's modulus of the piezoelectric material. For a piezoelectric island which
has large piezoelectric constants, such as Pb[Zr$_x{x}$Ti$_{1-x}$]O$_3$ (PZT) which has low absorption \cite{Pandey05}, the electric-field-induced strain can be as large as $\epsilon_e \sim 0.0002 \mbox{(MV/m)}^{-1}$, the corresponding Young's modulus is $Y\sim 10^5 \mbox{MPa}$ \cite{Zhang10}. The mediated coupling coefficient between the NV-spin (at a depth of 15 nm) and the electric field is thus $d\Delta/dE=(\epsilon_e\cdot Y)(d\Delta/d \sigma)=34\mbox{kHz}\cdot\mbox{(V/cm)}^{-1}$, as compared with the direct coupling coefficient of $\sim 10 \mbox{Hz}\cdot\mbox{(V/cm)}^{-1}$ \cite{Oort90}. The sensitivity for the measurement of electric field thus reaches $\eta_e=\eta_{\sigma}/(\epsilon_e\cdot Y)$. As the coupling between the NV center and the electric field is rather weak compared to the magnetic noise, and because the charge noise decreases rapidly with the distance from the NV center to the piezo-active layer \cite{Quant2002}, at a distance of e.g. $10-15$ nm, it is at least three orders of magnitude smaller than the magnetic noise effect considered above (see Supplementary). In the mean time, the charge noise in an isotopically engineered diamond \cite{Bala09} is also negligible. The sensitivity for the pressure measurement $\eta_{\sigma}\sim 0.35 \mbox{kPa} /\sqrt{\mbox{Hz}}$ therefore implies the sensitivity for the measurement of
electric field as high as $\eta_e\sim 0.2 (\mbox{V} \cdot \mbox{cm}^{-1}) /\sqrt{\mbox{Hz}}$, which
significantly exceeds the sensitivity of $202 (\mbox{V} \cdot \mbox{cm}^{-1}) /\sqrt{\mbox{Hz}}$ by a pure diamond NV sensor \cite{Dolde11}. This sensitivity would allow for the detection of the electric field produced by a single elementary charge at a distance from the NV-spin sensor of $\sim 8\mu m$ in around 1$s$, going beyond the single electron spin detection capacity of magnetic resonance force microscopy \cite{Rugar04},
and thus opens the possibility of remote sensing of a single charge at ambient conditions.  We remark that a similar sensitivity can be achieved in an NV-diamond electrically induced transparency device, which however requires an ensemble of NVs and liquid-nitrogen temperature \cite{Acosta13}.

In a similar way, it is possible to combine the hybrid diamond and piezomagnetic device with a thermally
sensitive element island on a substrate to measure temperature (see supplementary information). A change
of temperature induces the expansion of the thermally sensitive element island as characterized by its
thermal expansion constant $\epsilon_T$, which can be as high as $2.3\times 10^{-5} \mbox{K}^{-1}$
(aluminium). The thermal expansion produces a stress acting on the attached piezomagnetic layer as
$\sigma_T=(\epsilon_T \cdot \mathrm{T})\cdot Y$, where $Y$ denotes the Young's modulus of the thermally
sensitive material, e.g. $Y=7\times 10^4 \mbox{MPa}$ (aluminium). Thus, the response of the resonance frequency of an NV center at a depth of 15nm is $d\Delta/d T=(\epsilon_T\cdot Y)(d\Delta/d \sigma)=27\mbox{MHz}\cdot\mbox{K}^{-1}$, which represents significant improvement over the value $d\Delta/d T\sim 74\mbox{kHz}\cdot\mbox{K}^{-1}$ of a pure diamond NV sensor \cite{Acosta10prl}. The sensitivity for the measurement of temperature can reach $\eta_T=\eta_{\sigma}/(\epsilon_T\cdot Y) \sim 0.2 \mbox{mK} /\sqrt{\mbox{Hz}}$, and does not require an interrogation time of millisecond \cite{Kucs13,Toyli13,Neumann13}.

{\it Discussion and outlook.---} The proposed diamond-piezomagnetic hybrid sensor combines high
sensitivity under ambient conditions with nanometer scale dimensions. Such a device can find a
wide range of applications for measuring weak forces and pressures, tiny electric field and
temperature changes in many areas at ambient conditions, such as protein folding and DNA stretching in life science \cite{Pavone13}. Furthermore, it has implications for the study of fundamental
quantum physics phenomena such as the Casimir effect \cite{GarciaSanchezFB+12} or non-Newtonian
forces where forces have to be measured at smallest distances \cite{SushkovKD+11}. It may also
find applications in spin and mechanical oscillator hybrid systems. A hybrid diamond-piezomagnetic
(film) mechanical nanoresonator can significantly enhance the coupling between its vibrational
mode and NV electronic spins in diamond \cite{Bennett13,Albrecht13}, and thereby boost the phonon-mediated
effective spin-spin interactions. This would facilitate the spin squeezing at room temperature,
which can be used a resource for magnetometry and phonon-mediated quantum information processing.
The extension of the present model to arrays may find applications in tactile imaging \cite{Wang13}
and instantaneous pressure visualization by interactive electronic skin \cite{Javey13}. The proposed
device may also find applications in nanoscale surface characterization for example in read/write
heads of harddisks where the enhanced sensitivity to magnetic fields in combination with nanoscale
spatial resolution may significantly increase the characterisation speed which is a crucial constraint
in industrial applications.

{\it Acknowledgements.---} We thank Liam McGuinness and Boris Naydenov for discussion about experiments. This work was supported by the Alexander von Humboldt Foundation, the DFG
(FOR1482, FOR1493, SPP1601, and SFB TR21), the EU Integrating Projects SIQS and the EU STREPs EQUAM
and DIAMANT, DARPA via QUASAR, and the ERC Synergy grant BioQ. J.-M.C was supported also by a
Marie-Curie Intra-European Fellowship (FP7). Computations were performed on the bwGRiD.

\newpage 

\noindent\\
{\bf Author contributions}\\
J.M.C and M.B.P developed the concepts, J.M.C carried
out the theoretical calculations, all authors discussed
the work, J.M.C and M.B.P drafted the manuscript and
all authors commented on it.\\

\noindent\\
{\bf Additional information}\\
Supplementary information is available. Correspondence
and requests for materials should be addressed to J.M.C.
and M.B.P.\\

\onecolumngrid

\section*{Supplementary Information}

\setcounter{figure}{0}
\setcounter{equation}{0}

\noindent

{\it Micromagnetic simulation.---} For the calculation of the magnetization dynamics of the piezomagnetic material we
use the micromagnetic simulation as described by the Landau-Lifshitz-Gilbert (LLG) equation \cite{Landau35-SI,Gilbert55-SI}
as
\begin{equation}
    \frac{d\mathbf{M}_{k}}{d\tau}=-\gamma \mathbf{M}_{k} \times \mathbf{H}_{k} -\frac{\gamma
    \alpha}{M_S}\mathbf{M}_{k} \times\bla{\mathbf{M}_{k} \times \mathbf{H}_{k} }\label{eq:LL}
\end{equation}
where $\gamma$ is the gyro-magnetic ratio, $\alpha$ is the Gilbert damping constant, $\mathbf{M}_{k}=M_S \mathbf{m}_{k}$ is the magnetization direction in a single computation cell (domain) with $M_S $ the saturation magnetization of the material, and the unit vector $\mathbf{m}_{\alpha}=(\beta_1,\beta_2,\beta_3)$ with $\beta_i (i=1,2,3)$ being the direction cosines with respect to the crystal axes; $\mathbf{H}_{k} $ is the effective magnetic field as defined in the following equation by the variational derivation of the system energy $\mathbf{E}$ with respect to the magnetization
\begin{equation}
\mathbf{H}_{k} =-\frac{1}{\mu_0}\frac{\partial \mathbf{E}}{\partial \mathbf{M}_{k}},
\end{equation}
where $\mu_0$ is the magnetic permeability of the vacuum. The system energy $\mathbf{E}$ consists of the exchange energy, the magnetocrystalline anisotropy energy, the magnetostatic energy, the magnetoelastic energy, and the Zeeman energy \cite{Cullity72-SI}. In Eq.\ref{eq:LL}, the first term accounts for the gyromagnetic precession of the magnetization vector $\mathbf{M}_{k} $, and the second term arises from the viscous damping. Under a uniaxial stress along the direction $(\theta_1, \theta_2,\theta_3)$, where $\theta_i$ are the cosines of the angles between the direction vector and the three coordinate axes, the magnetoelastic energy $\mathbf{E}_{me}$ explicitly depends on the applied stress as follows \cite{Cullity72-SI}
\begin{equation}
     \mathbf{E}_{me}=\sigma\int \{-\frac{3}{2}\lambda_{100}\sum_{i=1}^3 \beta_i^2 \theta_i^2-
    \frac{3}{2}\lambda_{111}\!\!\sum_{i,j=1; i\neq j}^3 \beta_i \beta_j\theta_i\theta_j  \}d^3 r \label{eq:me}
\end{equation}
where $\lambda_{100}$ and $\lambda_{111}$ are the magnetostriction coefficients of the material along the $\langle 100 \rangle$ and $\langle 111 \rangle$ directions. The magnetoelastic energy $\mathbf{E}_{me}$ in Eq.\ref{eq:me} shows how an applied stress will change the magnetic behavior of the material. For example, a stress along the $\langle 111 \rangle$ direction tends to minimize the magnetoelastic energy by aligning the magnetization along the $\langle 111 \rangle$ direction. To achieve convergence of the numerical solution of micromagnetic dynamics, we choose the length of a single computation cell as $\Delta x=3.75\mbox{nm}$, which is smaller than the exchange length of Terfenol-D of $l_{exch}=\sqrt{2A/\mu_0 M_S^2} \sim 4.7 \mbox{nm}$, where $A$ is the exchange constant \cite{Rave98-SI}.  The parameters we use in our simulation are listed as follows:
\begin{table}[tbh]
\centering
{\large Tabel I: The parameters of Terfenol-D \cite{Clark86,Lord94,Lord97,Eng00}.}\\ \label{tab:parameters}
\begin{tabular}{ C{1in} C{1in} C{1.5in} C{1.5in}}\toprule[0.5pt]
 $\quad  M_S $& $\quad A$ &$\qquad (K_1,K_2)$  & $ \qquad (\lambda_{100},\lambda_{111}) $\\ \hline\hline
  $8$$\cdot 10^5$ A/m    & $9$$\cdot10^{-12}\mbox{J}/\mbox{m} $ & $(-0.8,-1.8) 10^5  \mbox{J}/\mbox{m}^3$     &  $\qquad (9,164) 10^{-5}$   \\
\bottomrule[0.5pt]
\end {tabular}\par
\label{table:tfd}
\end{table}

To account for the thermal fluctuation at a finite temperature $T$, we add a random thermal fluctuation field $\mathbf{H}_{th}$ to the effective magnetic field $\mathbf{H}_{k}$ in the micromagnetic dynamic equation (Eq.\ref{eq:me}), and describe the thermal field $\mathbf{H}_{th}$ as a Gaussian random process with the strength determined by the fluctuation-dissipation theorem \cite{Brown63-SI,Brown79-SI,Lazaro98-SI} as follows
\begin{equation}
\mathbf{H}_{th}^{m,i}(t)=\kappa^{m,i}(t) \bla{\frac{2\alpha k_B T}{\gamma\mu_0 M_S \Delta x^3\Delta t}}^{1/2},
\end{equation}
where $\kappa^{m,i}(t)$ is a normal distributed Gaussian random number that is uncorrelated in time and space while $k_B$ is the Boltzmann constant, $\Delta x$ is the length scale of the computation cell, and $\Delta t$ is the simulation time step which is chosen to be small enough in order to guarantee the convergence of the simulation results.\\

{\it Open system dynamics of NV-center spin.---} Taking into account the magnetic noise, the total Hamiltonian of the NV-center spin can be written as
\begin{equation}
H_t=H + \delta_{x}(t)S_x+\delta_{y}(t)S_y+\delta_{z}(t)S_z \label{eq-si:NV_Ham}
\end{equation}
where $H=D S_z^2+E(S_x^2-S_y^2)+\gamma \bla{\bar{B}_{x}S_x+\bar{B}_{y}S_y+ \bar{B}_{z}S_z}$, $\bar{B}_{\zeta}$ (with $\zeta=x,y,z$), is the mean value of the magnetic field acting on the NV-center spin along the $\hat{x},\hat{y},\hat{z}$ direction, $\delta_{\zeta}(t)=B_{\zeta}(t) -\bar{B}_{\zeta}$, represent the corresponding magnetic field fluctuation. The time correlation function of the magnetic noise is defined as $R_{\zeta}(t)=\langle \delta_{\zeta}(t) \delta_{\zeta}(0)\rangle$ for the magnetic noise arising from the magnetic material. As we find from our numeric simulations, the magnetic noise strength is weak and the correlation time scale is short enough (as compared to the coherence time of the NV-center spin), thus allowing us to make the Born-Markov approximation, and derive the following master equation to describe the decoherence dynamics of the NV-center spin under the influence of magnetic noise
\begin{equation}
    \frac{d\rho}{dt}=-i[H,\rho]+\mathcal{S}_{z}(0) \mathcal{L}^{\rho}(s_z)+\sum_{\kappa=\pm 1}
    \mathcal{S}_{\perp}(\omega_{\kappa})\mathcal{L}^{\rho}(s_{0\kappa})
\label{eq:qme}
\end{equation}
where $\mathcal{S}_{\perp}(\omega)=\blb{\mathcal{S}_{x}(\omega)+\mathcal{S}_{y}(\omega)}/2$ with $\mathcal{S}_{\zeta}(\omega)$ ($\zeta=x,y,z$) the magnetic noise power spectra, $\mathcal{L}^{\rho}(s_z)=s_z \rho s_z-\rho$ with $s_z=\ketbra{{+1}}{{+1}}-\ketbra{{-1}}{{-1}}$, $\mathcal{L}^{\rho}(s_{0\kappa})=L^{\rho}(\ketbra{{\kappa}}{{0}})+L^{\rho}(\ketbra{{0}}{{\kappa}})$ with $L^{\rho}(A)=A\rho A^{\dagger}-(1/2)(A^{\dagger} A \rho+\rho A^{\dagger}A)$.
The magnetic noise in parallel to the NV axis causes the dephasing of the NV-center spin, and the non-axial magnetic noise leads to relaxation.

\begin{figure}[h]
\begin{center}
\includegraphics[width=10cm]{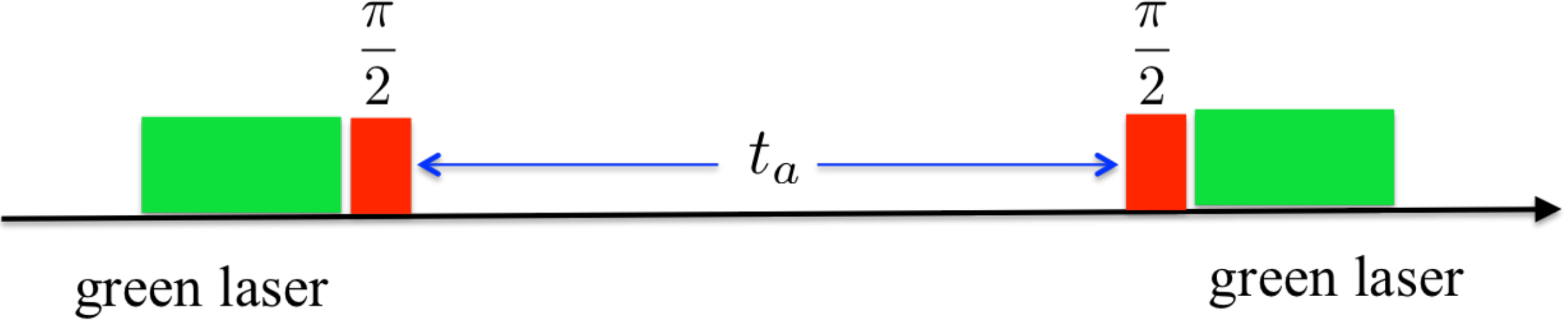}
\end{center}
\caption{The pulse sequence that is used to measure the effect of various external signals on the NV-center spin energy level structure. The NV-center spin is initialized by optical pumping with a green laser; which is further prepared into a coherent superposition of $\ket{{-1}}$ and $\ket{{+1}}$ with a microwave $(\pi/2)$ pulse. After a free evolution time $t_a$, the phase information resulting from external signals is mapped back to the sspin state population with another $(\pi/2)$ pulse, which is then readout by optical detection with a green laser.
}\label{fig:pulse_scheme}
\end{figure}

In Fig.\ref{fig:pulse_scheme}, we give a sketch of the pulse sequence utilized to measure the response of the NV-center spin to the external parameters that would affect its spin energy levels. The NV-center spin is first prepared into a coherent superposition state $\ket{\phi}=\sqrt{\frac{1}{2}}\bla{\ket{{-1}}+\ket{{+1}}}$ by applying a $(\pi/2)$ pulse with a microwave driving field $H_d=\Omega\blb{ \cos(\omega_{+1}  t)\ketbra{{+1}}{0}+\cos(\omega_{-1} t)\ketbra{{-1}}{0}}$ for a pulse duration $t_{\pi/2}=\pi/\Omega $.
From the master equation in Eq.(\ref{eq:qme}), we can write down the differential equations for the  density matrix elements of the NV-center spin state after a free evolution time $t_a$ (acquisition time), the solution of which is obtained as in Eq.(\ref{eq:sol_me})
\begin{equation}
    \rho(t)=\left( \begin{array}{ccc}
    p_{-1} & 0 &  q\\
    0 &  p_{0}& 0 \\
    q^{*} & 0 & \frac{1}{2} \end{array} \right), \label{eq:sol_me}
\end{equation}
where $p_{-1}=(1/4)\bla{1+\chi_{\perp}^2(t)}$, $p_0=(1/4)\bla{1-\chi_{\perp}^2(t)}$, $q=(1/2)\chi_{\perp}(t) \chi_{\parallel}(t) e^{-i\Delta t}$, and the relaxation coefficient is $\chi_{\perp}(t)=\exp{(-t \mathcal{S}_{\perp}(\omega_{-1})/2)}$. As the noise $\delta_{z}(t)$ can be assumed to be Gaussian, the dephasing coefficient can be calculated as follows
\begin{eqnarray}
 \chi_{\parallel}(t)&=&\langle \exp{\blb{i2\int_0^t \delta_{z}(s)ds}} \rangle  \\
&=& \exp{\blb{-\frac{1}{2}\int_0^t dt_1 \int_0^t dt_2 \langle 2 \delta_{z}(t_1)\cdot 2 \delta_{z}(t_2)\rangle}}\\
&=& \exp{\blb{-4 \int_0^t dt_1 \int_0^{t_1} dt_2 \langle  \delta_{z}(t_1) \cdot  \delta_{z}(t_2) \rangle}}\\
&=& \exp{\blb{-4 \int_0^t (t-s) R_{z}(s) ds}}\\
&=&\exp\blb{-4\int_{0}^{t} (t-s)R_{z}(0)e^{-\xi s} \cos(\omega_0 s) ds} \\
&\simeq&\exp\blb{-4 R_{z}(0)\frac{\xi }{\xi^2+\omega_0^2} t} \quad (\mbox{for} \quad \xi t\gg1 )\\
&\equiv&\exp{(-4t \mathcal{S}_{z}(0))}
\end{eqnarray}
The population of the NV-center spin state $m_s=0$ after a $(\pi/2)$ pulse, see Fig.\ref{fig:pulse_scheme}, is given by
\begin{equation}
P(\sigma,t_a)=\frac{1}{4}+\frac{p_{-1}}{2}+\frac{1}{2}\bla{q+q^*}
\end{equation}
which leads to the result of Eq.(3) in the main text.\\

\begin{figure}[h]
\begin{center}
\begin{minipage}{18cm}
\hspace{0cm}
\includegraphics[width=6.5cm]{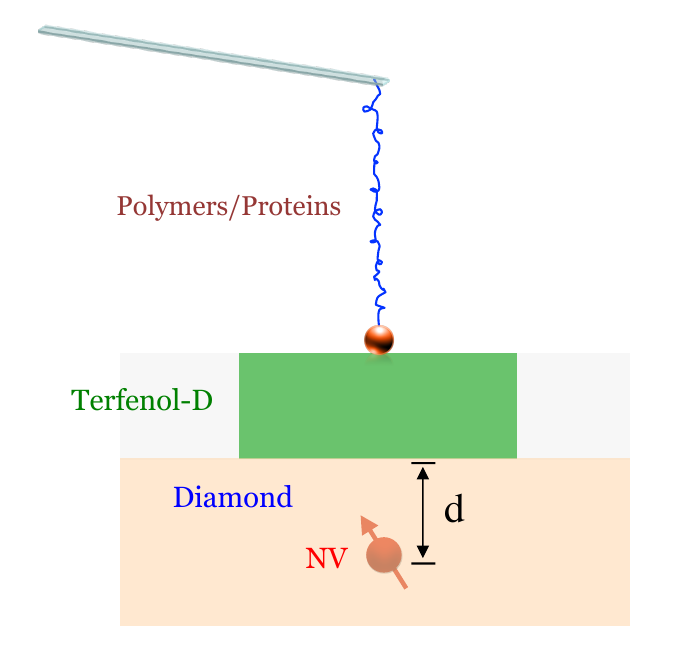}
\hspace{1cm}
\includegraphics[width=7cm]{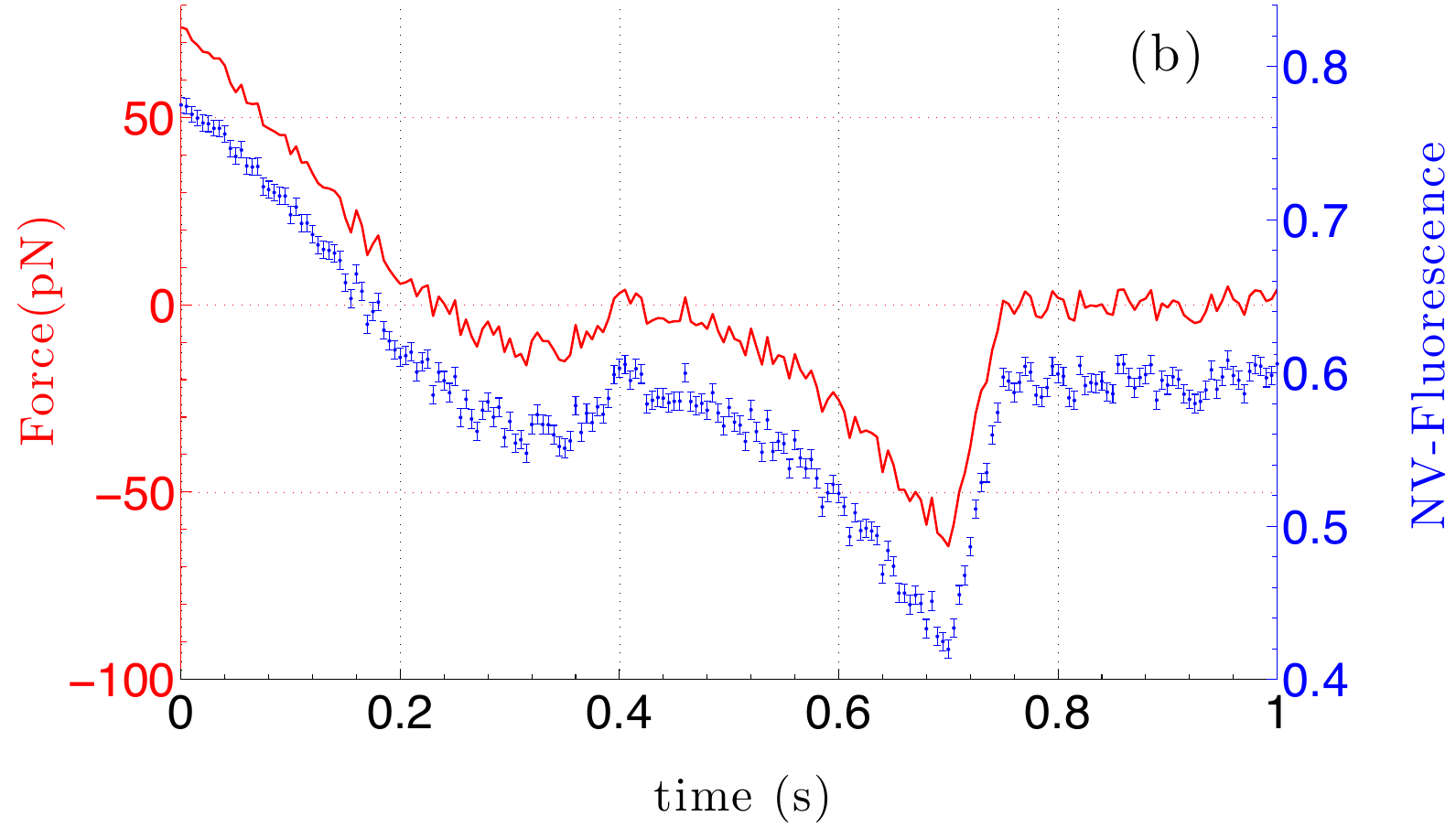}
\end{minipage}
\end{center}
\caption{Sketch of single-molecule force measurement. (a) Single molecules are attached with a cantilever and the surface of the hybrid diamond-piezomagnetic force sensor. (b) Measurement of DNA unbinding forces: The red curve shows a typical force-versus-time curve for the DNA duplex \cite{Strunz99-SI}, when the pulling speed is $v=100\mbox{nm/s}$. At a displacement of $\sim$50 nm (at time=0.7 s), the duplex unbinds at a loading force of $\sim$50 pN. The NV fluorescence curve (blue) is shifted downward by 0.05 for visibility. The system parameters are the same as Fig.3 of the main text.}\label{fig:force_meas}
\end{figure}

{\it Single-molecule force measurement.---} As a potential application of the diamond-piezomagnetic hybrid sensor, here we discuss its applicability to single molecule force spectroscopy experiments \cite{Neuman08-SI}. As a proof-of-principle, we calculate the NV center fluorescence signal when a single DNA molecule attached to the hybrid sensor is pulled to induce the dynamic unbinding force \cite{Strunz99-SI}. In Fig.\ref{fig:force_meas}, we consider a pulling speed of $v=100\mbox{nm/s}$. The spatial resolution is chosen as $\delta_d=0.5\mbox{nm}$ while the corresponding time resolution of $\tau_r=5$ ms is much longer than the nanosecond relaxation time scale of the piezomagnetic material. Therefore the magnetic material has sufficient time to respond to the change of force and can be assumed to be in equilibrium. We calculate the NV center fluorescence signal $P(t_a)$ with the acquisition time $t_a=150$ ns, and the corresponding shot noise is given by $\langle \Delta^2 P\rangle /\sqrt{\tau_r/(t_a+t_p)}$ (see the error bar in Fig.\ref{fig:force_meas}b), where the auxiliary time for the NV-center spin preparation and readout in each experiment run is set as $t_p=600$ ns. It can be seen from the result in Fig.\ref{fig:force_meas} that the detailed change of the (unbinding) force can be well observed. We anticipate that the approach can also be applied to other biological processes driven by molecular scale forces \cite{Neuman08-SI} and provides a good time and spatial resolution.\\

\begin{figure}[h]
\begin{center}
\hspace{2cm}
\includegraphics[width=6.5cm]{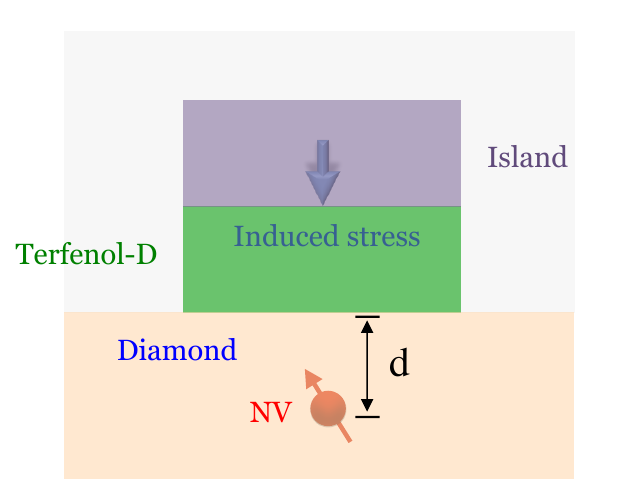}
\end{center}
\caption{A hybrid device that consists of a synthetic diamond layer formed by chemical vapour deposition (CVD) doped with NV centres, a piezomagnetic element layer, and a piezoelectric (thermally sensitive) element island encapsulated in a substrate. }\label{fig:electric_field_temperature}
\end{figure}

{\it Experiment setup for electric field and temperature measurement.---} The hybrid diamond-piezomagnetic pressure sensor can be used as a transducer to measure tiny electric field and temperature changes by incorporating with an additional island encapsulated in the substrate, see Fig.\ref{fig:electric_field_temperature}. If the island is made up of piezoelectric element, an internal mechanical strain in the piezoelectric element island is generated under the application of an electrical field. The electric-field-induced strain produces a force which is transcended to the piezomagnetic element layer, and is then detected by the hybrid diamond-piezomagnetic sensor. In a similar way, if we instead use the thermally sensitive element for the encapsulated island, a change in temperature leads to the thermal expansion of the island and thereby induces a stress acting on the piezomagnetic element layer, which can also be detected by the hybrid diamond-piezomagnetic sensor.

\end{document}